\documentclass[twocolumn]{jpsj2}

\def \d{{\rm d}}

\usepackage{amsmath}

\title{%
Role of Interlayer Electron Hopping  for 
 Spin Density Wave State \\
 in  the Zero-Gap  Organic Conductor 
}

\author{
Shinya KATAYAMA\thanks{E-mail address: katashin@slab.phys.nagoya-u.ac.jp}, 
Akito KOBAYASHI$^{1}$ and  Yoshikazu SUZUMURA 
}

\inst{
Department of Physics, Nagoya University, Nagoya 464-8602\\
$^{1}$Institute for Advanced Research, Nagoya University, Nagoya 464-8602
}

\recdate{August 23, 2007}

\abst{
  We investigate the formation of  density waves   in 
  the zero-gap state (ZGS)  which has been found 
 in the quasi-two-dimensional organic conductor $\alpha$-(BEDT-TTF)$_2$I$_3$ 
   salt  under the hydrostatic pressure.
 The  ZGS exhibits   the cone-like dispersion for 
  both  the conduction band and the valence band  which  
  degenerate each  other at  the two-dimensional 
  wave vectors, $\pm\mib{k}_0$  forming a zero gap.   
 By using the extended Hubbard model with repulsive interaction 
   we calculate the onset temperature of the spin density wave (SDW)  
    as a function of the interlayer electron hopping, which by itself 
    does not break the cone-like dispersion. 
It is shown that the SDW with wave number $2\mib{k}_0$ is induced 
  by the combined effect of the interaction,  the inter-band excitation  
  across the zero-gap and the interlayer hopping. 
}

\kword{$\alpha$-(BEDT-TTF)$_2$I$_3$, 
zero-gap state, interlayer electron hopping, spin density wave, 
hydrostatic pressure
}

\newcommand{\eti}{$\alpha$-(BEDT-TTF)$_2$I$_3~$}

\newcommand{\vep}{\varepsilon}
\newcommand{\lrangle}[1]{\langle{#1}\rangle}

\newcommand{\rmi}{\textrm{i}}

\newcommand{\jpsj}[3]{J. Phys. Soc. Jpn. {\textbf{#1}} ({#2}) {#3}}
\newcommand{\prb}[3]{Phys. Rev. B \textbf{#1} ({#2}) {#3}}
\newcommand{\prl}[3]{Phys. Rev. Lett. \textbf{#1} ({#2}) {#3}}

\begin{document}
\maketitle

%%%%%%%%%%%%%%%%%%%%%%%
\section{Introduction}\label{intro}
Organic conductors, which 
 have been investigated extensively for many years\cite{seo_chem_rev}, 
display various  states due to electronic correlation,  
for example, superconductivity, 
 Mott insulator or charge ordering 
 under varying  temperature or pressure. 

Recently, 
 it has been found that 
%  based on  the tight-binding calculation, 
  the 3/4-filled quasi-two-dimensional organic conductor 
   \eti salt  
 under  the uniaxial pressure along the BEDT-TTF molecule stacks ($a$-axis) 
  \cite{tajima_2002}
    exhibits  
      the novel electronic state described by the zero-gap state (ZGS).
\cite{katayama_zgs} 
Such an exotic state appears when  the Fermi surface is reduced to  a point, 
 and the valence and conduction bands degenerate 
  at two momenta ($\pm\mib{k}_0$) called contact point. 
  The linear dispersion around the contact point suggests 
      the  massless fermion.  
The existence of ZGS has been verified 
 by the first principle calculation\cite{ishibashi_1,kino_1}. 
The fact that the temperature dependence of the resistivity 
  becomes weak under the high pressure\cite{tajima_2000, tajima_2002} 
  has been  analyzed by using the Born approximation\cite{katayama_con}
  where  the life time of the electron is inversely proportional to the energy 
from the Fermi energy due to the linear density of states. 

 Although the ZGS is explained by the tight binding model, 
 there are also experimental evidences\cite{tajima_2002} showing  
   correlation effects in   \eti salt. 
The stripe charge ordering, which exists   perpendicular to the 
    stacking axis ($b$-axis) exists at ambient pressure and 
      under  low pressure\cite{takahashi}, 
%      ({\bf ref. of Takafashi's NMR})  
     has been analyzed  in terms of repulsive interactions 
      by the mean field theory
\cite{kino_co, seo_co,kobayashi_2004_sc}. 
 The origin of the superconductivity which  appears 
    with increasing  the uniaxial pressure along $a$-axis, $p_a$, 
   has been asserted to come from  the  spin fluctuation 
     in the presence of both the charge ordering and 
      the  Fermi surface.\cite{kobayashi_2005_sc}  
 At higher pressures, the superconductivity is suppressed 
   and  the  system becomes the ZGS. 
 %----------------------------------

The massless (or very light mass) 
   fermion system has been also found  
  in  graphite\cite{mcclure,slonczewski} 
   and bismuth\cite{wolff}, which 
  display various  properties, e. g. 
 anomalous diamagnetism\cite{mcclure,fukuyama_1970,kohno}, 
  the absence of backward scattering\cite{ando_backward,ando_review} and 
     the half-integer quantum hall effect\cite{novoselov,zhang}. 
However,  there are the following characteristics for 
the \eti salt\cite{katayama_zgs,kobayashi_zgs}
 compared with those of  graphite  and bismuth. 
The contact point of the \eti salt
  moves in the Brillouin zone under pressures.  
The Fermi velocity depends on the direction of  
  the momentum measured from the contact point.
 The effective Hamiltonian
 \cite{kobayashi_zgs} 
  consists of the Pauli 
  matrices $\sigma_x, \sigma_y, \sigma_z$ and $\sigma_0$
   on the bases of the wave functions 
    for the conduction band and the valence band at $\mib{k}_0$ 
   (or  $ - \mib{k}_0$)
   while  that of the  graphite
     \cite{mcclure}  
    consists of only $\sigma_x$ and  $\sigma_y$ on the basis of site representation.
%The equation corresponding to such an  effective Hamiltonian 
% is called as the Tilted Weyl equation.
%The effective Hamiltonian has been constructed 
% by the linear-expansion of $\mib{k}$ 
%  for the extended Hubbard Hamiltonian 
% at the contact point $\mib{k}_0$, 
 The  basis of these  effective Hamiltonian 
 %   the $\mib{k} \cdot \mib{p}$ expansion 
 %  on 
 is the Luttinger-Kohn representation
   \cite{luttinger}
   where wave functions are represented by the Bloch wave functions 
   at $\mib{k}_0$ (or  $ - \mib{k}_0$).
Further the interlayer electron hopping
\cite{kino_1} suggests new states 
 in the ZGS of  the present salt.
 %  based on    these profiles of the  \eti salt.

The ZGS state of the \eti salt under the hydrostatic pressure   
\cite{tajima_2006} 
exhibits 
 the anomalous behavior in the resistivity  at low temperature.
  With decreasing temperature,  
 there  is the rise of the resistance
 for  $T\lesssim10$ K.
The temperature dependence of the resistivity is strongly 
 influenced by the magnetic field, which is applied 
  to the $c$-axis being  perpendicular to the conducting plane. 
  For $H$  above 1 T,  
 the magnetoresistance 
   increases again at lower temperature ($T\lesssim10$ K)
 after showing a constant behavior  at the intermediate region 
 of temperature. 
 %--------------
For graphite and bismuth, 
  the magnetoresistance has only a hump.  
\cite{kopelevich_graphite, kopelevich_bismuth}. 
Thus the increase of the resistance indicates  
  an instability of the zero gap,  
 which is the subject of the  present paper.

   As a  possible state expected in the ZGS,    
   we study   the spin density wave (SDW) state 
    under the hydrostatic pressure, 
     which can be stabilized by the interlayer electron hopping. 
   Applying  the mean field theory to the extended Hubbard model, 
    the transition temperature $T_C$ for the SDW state  
     is calculated 
  where 
 the transfer energies for the hydrostatic pressure  
   are estimated  from  data of the uniaxial strain 
 and  the lattice distortion under the hydrostatic pressure.  
In \S2, we give formulations for estimating the transfer energies 
 under hydrostatic pressure, the charge disproportionation and 
   the linearized gap equation for density waves. 
The onset temperature of the SDW state is calculated in \S3. 
Summary and discussions  are given  in \S4.

%
%%%%%%%%%%%%%%%%%%%%%%%%%%%%%%%%%%%%%%
\section{Formulation}\label{formulation}
\subsection{model}\label{formulation1}
%%%%%%%%%%%%%%%%%%%%%%%%%%%%%%%%%%%%%%
\begin{figure}[tbp]
\begin{center}
\includegraphics[width=8.0cm]{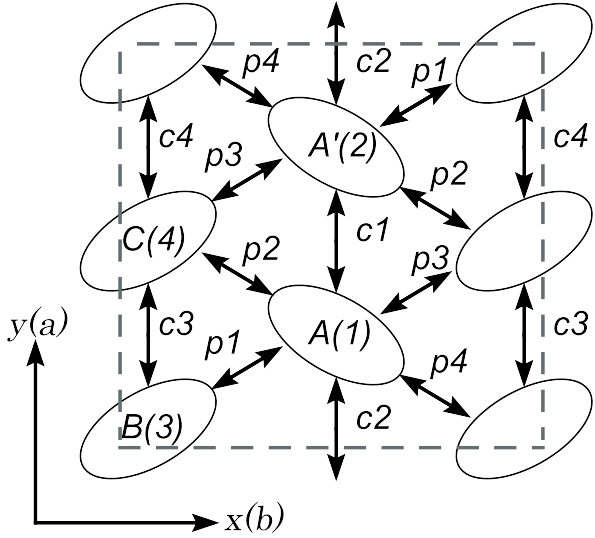}
\caption{Structure of the conducting plane of \eti 
where the unit cell is given by the dashed quadrangle, and  
$c1,\cdots,c4,p1,\cdots,p4$ denote the respective bonds 
 corresponding to the transfer energy of the electron hopping.
}
\label{unitcell}
\end{center}
\end{figure} 
%%%%%%%%%%%%%%%%%%%%%%%%%%%%%%%%%%%%%%
The crystal structure of the $\alpha$-type BEDT-TTF salt
  for the conducting  plane is shown in Fig. \ref{unitcell}, 
  where a unit cell consists of four BEDT-TTF molecules 
(A, A', B and C).
The $x,y$ axes
correspond to $a,b$ axes in the plane, respectively, while 
$z$ is the axis perpendicular to the plane.  
Transfer energies, which correspond to the intraplane  hopping 
and the interplane hopping, are respectively given by 
%%%%%%%%%%%%%%  (1) %%%%%%%%%%%%%%%%%%%%%%%%
 \begin{align}
 t_{i\alpha:j\beta} = t_A (A= c1, \cdots, p4) ,\;\;\; {\rm and} \;\;\; t_z \delta_{ij} \delta_{\alpha \beta}\; ,
\label{transfer_3d}
\end{align}
%%%%%%%%%%%%%%%%%%%%%%%%%%%%%%%%%%%%%%
 where $i,j$ are  the indices of the unit cell,  and   $\alpha$ and $\beta$ 
 correspond to A, A', B and C in a unit cell.  

The density wave is examined by  employing the  
extended Hubbard model,
%%%%%%%%%%%% (2)  %%%%%%%%%%%%%%%%%%%%%%%%%%
\begin{align}
H&=\sum_{n.n.,\sigma}
( t_{i\alpha:j\beta} 
a^{\dag}_{i\alpha\sigma}
a_{j\beta\sigma}+h.c.)\nonumber\\
&+
\sum_{i\alpha}U
a^{\dag}_{i\alpha\uparrow}
a^{\dag}_{i\alpha\downarrow}
a_{i\alpha\downarrow}
a_{i\alpha\uparrow}\nonumber\\
&+
\sum_{(n.n.),\sigma\sigma'}^{} V_{i\alpha:j\beta}
a^{\dag}_{i\alpha\sigma}
a^{\dag}_{j\beta\sigma'}
a_{j\beta\sigma'}
a_{i\alpha\sigma}
, 
\label{exh}
\end{align}
%%%%%%%%%%%%%%%%%%%%%%%%%%%%%%%%%%%%%%
where $a^{\dag}_{i\alpha\sigma}$, ($\sigma=\uparrow, \downarrow$) denotes a creation operator of the electron and 
  the first term is  the  hopping energy (eV). 
%   and $-(+)$ corresponds to spin with $\sigma = \uparrow (\downarrow)$.
 $n.n.$ denotes the nearest neighbor sites for both intraplane 
  and interplane. 
  The second and the third terms represent 
   the on-site and nearest neighbor repulsive interactions, respectively, 
   where $(n.n.)$ denotes the nearest neighbor site in  the intraplane. 
% with the length being unity. 
%The quantity $t_{i\alpha;j\beta}$ ($=t_A$ with $A=c1,\cdots,c4,p1,\cdots,p4$) 
%is the transfer energy between the $(j,\beta)$ site and 
%  the  $(i,\alpha)$ site for the intraplane,
%   and  $t_{i\alpha;j\beta} = t_z \delta_{\alpha \beta}$ is that 
%   for the interplane. 
%$U$ is the on-site Coulomb interaction, which is independent of the place of the molecule, 
The intersite interaction 
$V_{i\alpha;j\beta}(=V_c,V_p)$, which is considered  
only in a plane for the simplicity,   
 is taken as 
 $V_c$     for $c1,\cdots,c4$ bonds 
  and $V_p$ for $p1,\cdots,p4$ bonds.  
%  The unit of the energy is taken as eV.  
%The summation of the first and third term in eq. (\ref{exh}) takes 
%all bonds. 
% between nearest neighbor (i.e., n.n.) sites. 

%---------------------------------------
\subsection{transfer energy under hydrostatic pressures}\label{formulation2}

We estimate the intraplane transfer energy (Fig.~\ref{unitcell})
 under the hydrostatic pressure 
  by assuming the relation,  
%%%%%%%%%%%%%%% (3) %%%%%%%%%%%%%%%%%%%%%%%
\begin{align}
t_A  = t_A (0) (1 + K^a_A p_a+K^b_A p_b), 
\label{t_and_p_0}
\end{align}
%%%%%%%%%%%%%%%%%%%%%%%%%%%%%%%%%%%%%%
where $p_a(p_b)$ is the pressure along $a(b)$ direction
and  $t_A (0)$ corresponds to the energy at ambient pressure. 
Since $t_A $ is  not  known directly, 
 coefficients $K^a_A$ and $K^b_A$ are estimated 
 by using  the  data of the uniaxial strain. 
When  the uniaxial strain  $p_a$ ($p_b$) is applied along 
$a$($b$)-axis,  
 the pressure  $p_b' (p_a')$ along 
 the  $b$($a$)-axis  is also added, where 
%%%%%%%%%%% (4)  %%%%%%%%%%%%%%%%%%%%%%%%%%%
\begin{align}
 p_b'&=r_1 p_a, \nonumber \\
 p_a'&=r_2 p_b . 
\label{ratio_P}
\end{align} 
%%%%%%%%%%%%%%%%%%%%%%%%%%%%%%%%%%%%%%
 Such an additional pressure  
 is needed to  cancel  out the Poisson's effect, i.e., 
 to retain no lattice distortion along the  $b$($a$)-axis. 
%which makes the uniaxial stain of  $b$($a$) zero. 
Coefficients  $r_1$ and $r_2$ are estimated from 
the data of lattice parameters as   
$r_1 = 0.335$ and 
$r_2 = 0.257$, which are derived in Appendix. 

Using eqs.~(\ref{t_and_p_0}) and (\ref{ratio_P}), 
$K^a_A$ and $K^b_A$ are obtained as 
%can be solved from eq. (\ref{t_and_p}) under the uniaxial strain as 
%%%%%%%%%%%  (5) %%%%%%%%%%%%%%%%%%%%%%%%%%
\begin{align}
\label{eq:K-L}
\left(
\begin{array}{c}
K^a_A\\
K^b_A
\end{array}
\right)
&=
\dfrac{1}{1-r_1r_2}
\left(
\begin{array}{cc}
1&-r_1\\
-r_2&1
\end{array}
\right)
\left(
\begin{array}{c}
L^a_A\\
L^b_A
\end{array}
\right), 
\end{align}
where
%----------  (6) -----------
\begin{align}
L_A^a \equiv \frac{1}{t_A(0)} \cdot \frac{\d t_A(p_a)}{\d p_a }, \; \;
L_A^b \equiv  \frac{1}{t_A(0)} \cdot \frac{\d t_A(p_b)}{\d p_b }. 
\label{t_and_p}
\end{align}
 Quantities $L^a_A$ and $L^b_A$ denote 
  the variation of transfer energies 
 with respect to uniaxial strain  $p_a$ and $p_b$, respectively. 
Equation~(\ref{t_and_p}) is calculated    
 from the transfer energies under the 
 ambient pressure, 
$a$-axis strain at 2 kbar and $b$-axis strain at 3 kbar\cite{katayama_zgs,kondo}(see Table \ref{table_coefficient}).
Substituting $r_1$, $r_2$, $L_A^a$ and $L_A^b$  into eq.~(\ref{eq:K-L}),
 we obtain 
 the transfer energy, 
$ t_A (p) = t_A (0) (1 + (K^a_A + K^b_A) p)$ for the hydrostatic pressure, 
 where 
 $K^a_A+K^b_A$ are given in Table \ref{table_coefficient}.

%%%%%%%%%%%  Table 1 %%%%%%%%%%%%%%%%%%%%%%%%%%%
\begin{table*}[tbp]
\caption{ 
Transfer energy per unit pressure, which is normalized 
 by $t_A(0)$.
$L^a_A$ and $L^b_A$ are obtained in Ref. \citen{katayama_zgs} 
based on Ref. \citen{kondo}. 
}
\begin{tabular}{ccccccccc}
\hline
$A$&$c1$&$c2$&$c3$&$c4$&$p1$&$p2$&$p3$&$p4$\\
\hline
$a$-axis strain ($L^a_A$)
&0.167&-0.025&0.089&0.089&0.011&0.0&0.0&0.032\\
$b$-axis strain ($L^b_A$)
&0.042&0.133&0.167&0.167&0.024&0.022&0.053&0.032\\
Hydrostatic pressure ($K^a_A+K^b_A$)
&0.166&0.077&0.194&0.194&0.026&0.016&0.039&0.042\\
\hline
\end{tabular}
\label{table_coefficient}
\end{table*}
%%%%%%%%%%%%%%%%%%%%%%%%%%%%%%%%%%%%%%

%--------------------------------
\subsection{charge disproportionation}\label{formulation3}

Next, we  calculate  the density 
 $\lrangle{n_{\alpha}}
(=(1/N)\sum^{}_{i\sigma}\lrangle{a^{\dag}_{i\alpha\sigma}a_{i\alpha\sigma}})$ 
 by using the self-consistent 
 Hartree approximation, which gives the nonmagnetic state
  and the ZGS.
The charge disproportionation corresponds to   
  the spatial variation of $\lrangle{n_{\alpha}}$. 
 %  where there is no magnetic moment at every  site. 
In a way similar  to previous studies, 
\cite{kobayashi_zgs,seo_co,kobayashi_2004_sc,kobayashi_2005_sc}
the self-consistent equations for the density 
 at the respective site $\alpha$  
 are written as $(\mib{k} = (k_x, k_y, k_z))$
%%%%%%%%%%%% (7) %%%%%%%%%%%%%%%%%%%%%%%%%%
\begin{align}
%-----------------
&
\langle n_{\alpha} \rangle = 2 \sum_{\gamma=1}^{4}
|d_{\alpha\gamma}(\mib{k})|^2
%\ d^{}_{\alpha\gamma}(\mib{k})
\dfrac{1}{\exp{[\xi_{\gamma}(\mib{k})/T]}+1} , 
%-----------------
%\nonumber\\
&
%---------------------
%\qquad+
%\dfrac{1}{N}
%\sum_{i,j}
%t_{i\alpha;j\beta} 
%e^{{\rm i} \mib{k}\cdot (\mib{r}_i-\mib{r}_j)}, 
\label{hmf}
\end{align}
where 
%----------  (8) (9) (10) --------------
\begin{align}
&\sum_{\beta=1}^{4}
( \varepsilon_{\alpha \beta}^{}(\mib{k}) +
t^{}_{\alpha}\delta_{\alpha\beta}- \mu \delta_{\alpha\beta})\ 
d^{}_{\beta \gamma}(\mib{k})
=\xi_{\gamma}(\mib{k})\ d^{}_{\alpha \gamma}(\mib{k}) ,
\label{eigenvalue} 
\\
&
 \varepsilon_{\alpha \beta}^{}(\mib{k}) = \frac{1}{N}
  \sum_{ (n.n.)} t_{i \alpha : j \beta}
   {\rm e}^{- i \mib{k} (\mib{r}_i-\mib{r}_j )}
    + 2 t_z \cos k_z \delta_{\alpha \beta} \; ,
\label{eigenvalue_kin} 
%----------------------
\\
&
t^{}_{\alpha}
=\dfrac{U\langle n_{\alpha } \rangle}{2}
+\dfrac{1}{N}
\sum_{(n.n.)} 
V_{i\alpha;j\beta'} \langle n_{\beta'} \rangle 
 .
\label{eigensystem}
\end{align}
%%%%%%%%%%%%%%%%%%%%%%%%%%%%%%%%%%%%%%
 $N$ is the number of the unit cell. 
For eqs.~(\ref{hmf}) and (\ref{eigenvalue}), 
 we used the Fourier transformation expressed as 
$a_{i\alpha\sigma}=(1/\sqrt{N})
\sum_{\mib{k}}c_{\mib{k}\alpha\sigma}e^{-{\rm i}\mib{k}\cdot\mib{r}_i}$.  
 We take the lattice constant as unity.
The  band index $\gamma$ 
% in the eigenvalue $\xi_{\gamma\sigma}(\mib{k})$ and 
%   eigenvector $d_{\alpha\gamma}(\mib{k})$ 
 is taken as the descending order of $\xi_{\gamma}(\mib{k})$, i.e.,  
$\xi_{1}(\mib{k})>\xi_{2}(\mib{k})>\xi_{3}(\mib{k})>\xi_{4}(\mib{k})$. 
The quantity $\mu$ is the chemical potential determined by 
 a condition, 
$\sum^{}_{\alpha}\lrangle{n_{\alpha}}=6$, 
due to  3/4 filling. 
%  We  calculate eq.(\ref{hmf}) with $t_z=0$ since 
The effect  of $t_z$ on the charge disproportionation and $\mu$ 
  is negligibly small, e.g., 
  $\lrangle{\delta n_{\alpha}}\sim10^{-5}$ and $\delta\mu\sim10^{-5}$ eV
  in the present calculation of $t_z \sim 0.003$.  
%{\bf (How much do $\mu$ and $<n_{\alpha}>$ change by $t_z$?) }
The temperature ($T$)
%in eq. (\ref{eigensystem}) 
 is taken  as $T \rightarrow 0$ 
 in calculating the charge disproportionation, 
 since the $T$ dependence of  $\lrangle{n_{\alpha}}$
 is negligibly small 
   at  temperatures  for the onset of  SDW. 
The parameters  of  $U, V_p, V_c$ and $p$ are chosen 
 to obtain
   the ZGS state in the presence of charge disproportionation.
%    given by  
%    $\langle n_{\beta' \sigma'} \rangle$.  
% For the present calculation, there is no magnetic moment in the ZGS, i. e. 
%$\langle n_{\alpha\uparrow} \rangle=\langle n_{\alpha\downarrow} \rangle$
%for $H=0$. 

%--------------------------------------
\subsection{transition temperature for density waves}\label{formulation4}

%%%%%%%%%%%%  Fig 2 %%%%%%%%%%%%%%%%%%%%%%%%%%
\begin{figure}[tbp]
\begin{center}
\includegraphics[width=8.0cm]{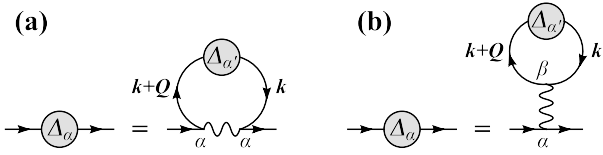} 
\caption{Feynman diagrams 
for $\Delta^{Sx}_{\mib{Q}\alpha}$ and $\Delta^{Sy}_{\mib{Q}\alpha}$ 
in eq. (\ref{deltaxy}) 
(a), and that 
for $\Delta^{Sz}_{\mib{Q}\alpha}$ in eq. (\ref{deltaxy}) and $\Delta^{C}_{\mib{Q}\alpha}$ 
in eq. (\ref{deltac}) (b).
}
\label{diagram}
\end{center}
\end{figure} 
%%%%%%%%%%%%%%%%%%%%%%%%%%%%%%%%%%%%%%
Now, we calculate the transition temperature, $T_C$, corresponding to 
the onset temperature for the density waves, where 
the operator of the spin density waves, 
($S^x_{\mib{Q}\alpha}(\mib{k}), S^y_{\mib{Q}\alpha}(\mib{k}), S^y_{\mib{Q}\alpha}(\mib{k})$) and 
 that of the  charge density wave ($C_{\mib{Q}\alpha}(\mib{k})$), 
 are defined, respectively,  as $(\mib{Q} =(q_x, q_y, q_z))$
%%%%%%%%%%% (11) %%%%%%%%%%%%%%%%%%%%%%%%%%%
\begin{align}
\label{eq:density_operator}
S^x_{\mib{Q}\alpha}(\mib{k})&=
(c^{\dag}_{\mib{k}+\mib{Q}\alpha\uparrow}c_{\mib{k}\alpha\downarrow}+
c^{\dag}_{\mib{k}+\mib{Q}\alpha\downarrow}c_{\mib{k}\alpha\uparrow})/2 \; , 
\nonumber\\
S^y_{\mib{Q}\alpha}(\mib{k})&=-\rmi
(c^{\dag}_{\mib{k}+\mib{Q}\alpha\uparrow}c_{\mib{k}\alpha\downarrow}-
c^{\dag}_{\mib{k}+\mib{Q}\alpha\downarrow}c_{\mib{k}\alpha\uparrow})/2 \; , 
\nonumber\\
S^z_{\mib{Q}\alpha}(\mib{k})&=
(c^{\dag}_{\mib{k}+\mib{Q}\alpha\uparrow}c_{\mib{k}\alpha\uparrow}-
c^{\dag}_{\mib{k}+\mib{Q}\alpha\downarrow}c_{\mib{k}\alpha\downarrow})/2 \; , 
\nonumber\\
C_{\mib{Q}\alpha}(\mib{k})&=
(c^{\dag}_{\mib{k}+\mib{Q}\alpha\uparrow}c_{\mib{k}\alpha\uparrow}+
c^{\dag}_{\mib{k}+\mib{Q}\alpha\downarrow}c_{\mib{k}\alpha\downarrow})/2 \; . 
\end{align}
%%%%%%%%%%%%%%%%%%%%%%%%%%%%%%%%%%%%%%
In terms of eq.~(\ref{eq:density_operator}), 
 the second and third terms of eq. (\ref{exh}) can be rewritten as  
%%%%%%%%%%% (12) %%%%%%%%%%%%%%%%%%%%%%%%%%%
\begin{align}
&\sum_{i\alpha}U
a^{\dag}_{i\alpha\uparrow}
a^{\dag}_{i\alpha\downarrow}
a_{i\alpha\downarrow}
a_{i\alpha\uparrow}\nonumber\\
&=-\dfrac{1}{N}
\sum_{
\mib{kk'Q}\alpha
}
U(S^{x}_{-\mib{Q}\alpha}(\mib{k}')S^{x}_{\mib{Q}\alpha}(\mib{k})+
S^{y}_{-\mib{Q}\alpha}(\mib{k}')S^{y}_{\mib{Q}\alpha}(\mib{k}))
\nonumber\\
&=
\dfrac{1}{N}
\sum_{
\mib{kk'Q}\alpha
}
U(C_{-\mib{Q}\alpha}(\mib{k}')C_{\mib{Q}\alpha}(\mib{k})-
S^{z}_{-\mib{Q}\alpha}(\mib{k}')S^{z}_{\mib{Q}\alpha}(\mib{k}))
\end{align}
%%%%%%%%%%%%%%%%%%%%%%%%%%%%%%%%%%%%%%
and 
%%%%%%%%%%%%%% (13) %%%%%%%%%%%%%%%%%%%%%%%%
\begin{align}
&\sum_{(n.n.),\sigma\sigma'}V_{i\alpha;j\beta}
a^{\dag}_{i\alpha\sigma}
a^{\dag}_{j\beta\sigma'}
a_{j\beta\sigma'}
a_{i\alpha\sigma}\nonumber\\
&=
\dfrac{1}{N}
\sum_{
\mib{kk'Q}\alpha\beta
}
2V_{\alpha\beta}(\mib{Q})
C_{-\mib{Q}\beta}(\mib{k}')C_{\mib{Q}\alpha}(\mib{k}),
\label{gapequation_sc}
\end{align}
%%%%%%%%%%%%%%%%%%%%%%%%%%%%%%%%%%%%%%
respectively.
The  matrix elements
of $V_{\alpha\beta}(\mib{Q})$ are given as  
%----------  (14) ---------------
\begin{align}
& V_{\alpha\alpha}(\mib{Q})=0, 
 \nonumber \\ &
V_{12}(\mib{Q})=V_c(1+e^{-\rmi q_y}),
 \nonumber \\ &
V_{13}(\mib{Q})=V_p(1+e^{\rmi q_x}),
 \nonumber \\ &
V_{14}(\mib{Q})=V_p(1+e^{\rmi q_x}),
 \nonumber \\ &
V_{23}(\mib{Q})=V_p(e^{\rmi q_x}+e^{\rmi (q_x+q_y)}),
 \nonumber \\ &
V_{24}(\mib{Q})=V_p(1+e^{\rmi q_x}),
 \nonumber \\ &
V_{34}(\mib{Q})=V_c(1+e^{-\rmi q_y}),
 \end{align} 
and $V_{\alpha\beta}(\mib{Q})=V^*_{\beta\alpha}(\mib{Q})$.
Applying the mean field approximation, eq. (\ref{exh}) is expressed as 
%%%%%%%%%%%%% (15) %%%%%%%%%%%%%%%%%%%%%%%%%
\begin{align}
H_{\rm MF}&=\sum_{\mib{k}\gamma\sigma}\xi_{\gamma}(\mib{k})
c^{\dag}_{\mib{k}\gamma\sigma}
c_{\mib{k}\gamma\sigma}
\nonumber\\
&+
\sum_{\mib{k}\mib{Q}\alpha}
\left[
\Delta^{S_x}_{\mib{Q}\alpha}
S^x_{\mib{Q}\alpha}(\mib{k})
 + 
\Delta^{S_y}_{\mib{Q}\alpha}
S^y_{\mib{Q}\alpha}(\mib{k})
+h.c.\right]
\label{eq_delta_SXY}
%+
%\Delta^{C}_{\mib{q}\alpha}
%C_{\mib{q}\alpha}(\mib{k})+h.c., 
%\nonumber 
\end{align}
or  
%--------- (16) -----------------------
\begin{align}
H_{\rm MF}&=\sum_{\mib{k}\gamma\sigma}\xi_{\gamma}(\mib{k})
c^{\dag}_{\mib{k}\gamma\sigma}
c_{\mib{k}\gamma\sigma}
\nonumber\\
&+
\sum_{\mib{k}\mib{Q}\alpha}
\left[
\Delta^{S_z}_{\mib{Q}\alpha}
S^z_{\mib{Q}\alpha}(\mib{k})
 +
\Delta^{C}_{\mib{Q}\alpha}
C_{\mib{Q}\alpha}(\mib{k})+h.c.\right], 
\label{eq_delta_SZ}
\end{align}
%%%%%%%%%%%%%%%%%%%%%%%%%%%%%%%%%%%%%%
where $\Delta^{S_{\zeta}}_{\mib{Q}\alpha}$, 
$(\zeta = x, y, z)$ and $\Delta^{C}_{\mib{Q}\alpha}$ are 
the order parameters of the SDW and CDW state 
defined as
%%%%%%%%% (17) %%%%%%%%%%%%%%%%%%%%%%%%%%%%%
\begin{align}
\Delta^{S_{\zeta}}_{\mib{Q}\alpha}&=-\dfrac{1}{N}
\sum^{}_{\mib{k}}U\lrangle{S^{\zeta}_{-\mib{Q}\alpha}(\mib{k})}\nonumber\\
\Delta^{C}_{\mib{Q}\alpha}&=\dfrac{1}{N}
\sum^{}_{\mib{k}\beta}(U\delta_{\alpha\beta}+2V_{\alpha\beta}(\mib{Q}))
\lrangle{C_{-\mib{Q}\beta}(\mib{k})}. 
\label{gap_equation}
\end{align}
%%%%%%%%%%%%%%%%%%%%%%%%%%%%%%%%%%%%%%
The quantity $\xi_{\gamma}(\mib{k})$ 
in eqs.~(\ref{eq_delta_SXY}) and (\ref{eq_delta_SZ})
  is the kinetic energy, 
which exhibits the cone-like dispersion 
 and the zero gap located between  
  $\xi_{1}(\mib{k})$  and  
   $\xi_{2}(\mib{k})$.
   % in  eq. (\ref{eigenvalue}).
 The quantity   
   $c_{\mib{k}\gamma\sigma}(=\sum^{}_{\alpha}d^{*}_{\alpha\gamma}(\mib{k})c_{\mib{k}\alpha\sigma})$ 
is the annihilation operator for the particle  
with the wave number $\mib{k}$  
 and spin  $\sigma$ in the band $\gamma$. 

 The transition temperature $T_C$ for the density wave is calculated 
 by using  the linearized gap equations for 
${\Delta}^{S_{\zeta}}_{\mib{Q}\alpha}$, $(\zeta = x, y, z)$ 
 and $\Delta^{C}_{\mib{Q}\alpha}$, 
 which are shown in Fig. \ref{diagram}.
%where the diagrams of eq. (\ref{gap_equation}) is drawn in Fig. \ref{diagram}.
They are written explicitly as 
%%%%%%%%%%% (18) (19) %%%%%%%%%%%%%%%%%%%%%%%%%%%
\begin{align}
\lambda^{S}
\Delta^{S\zeta}_{\mib{Q}\alpha}
=&
\sum_{\alpha'\gamma\gamma'}U
I_{\gamma\gamma'}(\mib{Q};\alpha,\alpha')
\Delta^{S\zeta}_{\mib{Q}\alpha'}
\label{deltaxy}
\\
%%%%%%%%%%%%%%%%%%
%\lambda^{Sz}
%\Delta^{Sz}_{\mib{q}\alpha}
%=&\sum_{\alpha'\gamma\gamma'}U
%I^{\perp}_{\gamma\gamma'}(\mib{q};\alpha,\alpha')
%\Delta^{Sz}_{\mib{q}\alpha'}
%\label{deltaz}
%\\
%%%%%%%%%%%%%%%%%%%%%%%%%%%%%%%%%%%%
\lambda^{C}
\Delta^{C}_{\mib{Q}\alpha}
=&-\sum_{\beta\alpha'\gamma\gamma'}
(U\delta_{\alpha\beta}+2V_{\alpha\beta}(\mib{Q}))\nonumber\\
&\times I_{\gamma\gamma'}(\mib{Q};\beta,\alpha')
\Delta^{C}_{\mib{Q}\alpha'}, 
\label{deltac}
\end{align}
%%%%%%%%%%%%%%%%%%%%%%%%%%%%%%%%%%%%%%
where $I_{\gamma\gamma'}(\mib{Q};\alpha,\beta)$ 
is defined as 
%and 
%$I^{\perp}_{\gamma\gamma'}(\mib{q};\alpha,\beta)$ 
%{\bf (what is the meaning of parallel and perpendicular) }
%%%%%%%%%%%%% (20) %%%%%%%%%%%%%%%%%%%%%%%%%
\begin{align}
\label{eq:density_response}
I_{\gamma\gamma'}(\mib{Q};\alpha,\beta)&=
-\dfrac{1}{N}\sum_{\mib{k}}
\dfrac{f(\xi_{\gamma}(\mib{k}+\mib{Q}))-
f(\xi_{\gamma'}(\mib{k}))}
{\xi_{\gamma}(\mib{k}+\mib{Q})-\xi_{\gamma'}(\mib{k})}
\nonumber\\
&\times
d^{}_{\alpha \gamma}(\mib{k}+\mib{Q})
d^{*}_{\alpha \gamma'}(\mib{k})
d^{*}_{\beta \gamma}(\mib{k}+\mib{Q})
d^{}_{\beta \gamma'}(\mib{k})
\end{align}
%%%%%%%%%%%%%%%%%%%%%%%%%%%%%%%%%%%%%%
and $f(x)(=[e^{x/T}+1]^{-1})$ is the Fermi distribution function. 
The transition temperature of the density wave 
is obtained from the condition $\lambda^S=1$ or $\lambda^C=1$.
We note that spin response parallel to the quantized axis 
 ($z$-axis) 
 is the same as that of perpendicular to the $z$-axis 
 due to isotropic properties for both interactions and 
  transfer energies. 
%In the present study, the wave number of the density wave takes only 
%$\mib{q}=0$ or $\mib{q}=2\mib{k}_0$ since the energy spectrum without 
%the ordered state shows the zero-gap state, 
%which the contact point exists at $\pm\mib{k}_0$. 
The summation of $\gamma$ and $\gamma'$ in 
eqs.~(\ref{deltaxy}) and (\ref{deltac}) 
takes only the conduction band ($\gamma =1$)
and the valence band ($\gamma =2$) 
since the density wave in the present calculation 
is determined essentially 
by these two bands. 
%In calculating $\lambda$ in eqs. \ref{deltaxy}, \ref{deltaz}, or \ref{deltac}, 
%the summation of $\mib{k}'$ does not take only uniformly in the first Brillouin zone but also 
%take minutely in the region $-0.3\pi< \pm k^{x,y}_0 < 0.3\pi$. 
Hereafter, the units of the pressure and energy are 
  taken  as kbar and eV, respectively. 
We also use  the two-dimensional component   
 of   $\mib{Q}$, i.e.,  
%$\mib{q}$ as  
%------------ (21) -------------- 
\begin{eqnarray}
 \mib{Q} \equiv \mib{q} +  q_z \mib{\rm e}_z ,
\end{eqnarray}
where $\mib{q}=(q_x,q_y)$ and  $\mib{\rm e}_z$ is the unit vector 
perpendicular to the conducting plane.

%%%%%%%%%%%%%%%%%%%%%%%%%%%%%%%%%%%%%%
\section{Spin Density Wave}
%The numerical calculation is performed by setting  parameters 
%  as $U=0.4$, $V_c=0.17$, $V_p=0.05$ and $p=12$. 
Before studying the SDW state, 
 we describe the state  where the interlayer transfer is absent. 
  The ZGS  under the hydrostatic pressure is found as follows. 
When only the kinetic energy is taken into account, i.e., 
  without the Coulomb interaction, 
   the Fermi surface   exists 
   for $0<p<5.4$ where 
     the contact point is located below  the Fermi energy. 
With increasing the pressure, 
 the Fermi surface becomes small and the ZGS 
     is obtained for $p > 5.4$. 
The introduction of interactions  suppresses the ZGS state.
 When  we take a set of  parameters,  $U=0.4, V_p=0.05, V_c=0.17$
 \cite{kobayashi_2005_sc}
  as is used in the present calculation,  
  there exists the insulating state with  the stripe charge order 
     for $0<p<2.1$,  the  metallic state with the  
      charge order for  $2.1<p<12.5$,
%      {\bf (Is this CO the same as that of low pressure?)}
       and 
      finally the ZGS for  $p>12.5$.

%%%%%%%%%%%%%%  Fig. 3    %%%%%%%%%%%%%%%%%%%%%%%%
\begin{figure}[tbp]
\begin{center}
\includegraphics[width=8.0cm]{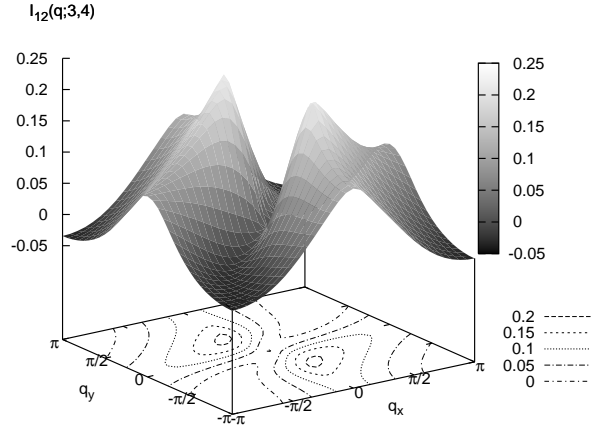}
\caption{$\mib{q}$-dependence of 
$I_{12}(\mib{q};\alpha,\beta)\ ((\alpha,\beta)=(3,4))$ 
for 
$U=0.4$, $V_c=0.17$, $V_p=0.05$ and $p=16$, where the 
contact points exist on $\mib{k}_0=(\pm0.970\pi,\pm0.216\pi)$. 
The temperature is chosen as $T=0.001$. 
}
\label{chiq}
\end{center}
\end{figure} 
%%%%%%%%%%% Fig. 4  %%%%%%%%%%%%%%%%%%%%%%%%%%%
\begin{figure}[tbp]
\begin{center}
\includegraphics[width=7.0cm]{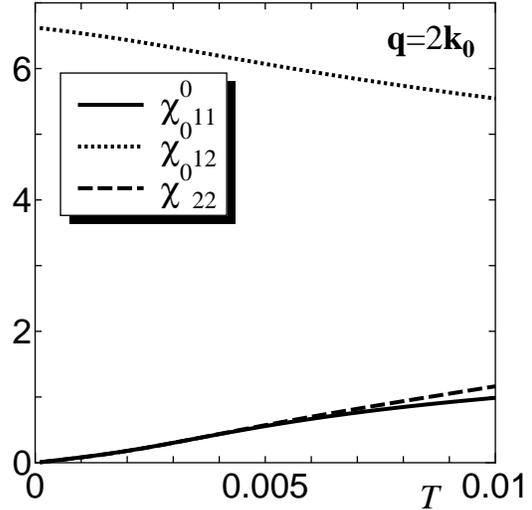}
\caption{
$T$-dependence of the bare susceptibility, 
$\chi^{0}_{\gamma\gamma'}(\mib{q})$ for 
$U=0.4$, $V_c=0.17$, $V_p=0.05$ and $p=16$. 
The cases for $(\gamma,\gamma')=(1,1)$, $(2,2)$ and $(1,2)$ are plotted by 
solid, dashed and dotted line, respectively. 
The (2,1) component of the susceptibility is equal to the (1,2) component. 
The wave number is taken as $\mib{q}=2\mib{k}_0(=(-0.060\pi, 0.432\pi))$. }
\label{chi0}
\end{center}
\end{figure} 
%%%%%%%%%%%%%%%%%%%%%%%%%%%%%%%%%%%%%%

%----------------------------------

\subsection{density waves  in the absence of $t_z$}
We examine 
the  property of $I_{\gamma\gamma'}(\mib{q};\alpha,\beta)$ 
 for $t_z=0$. 
 The density response function of eq.~(\ref{eq:density_response})
 consists  of the intra-band components $I_{11}$ and $I_{22}$ and 
  the inter-band component $I_{12} (=I_{21})$ where 
 the main contribution comes from  the inter-band one 
  having the effect similar to the nesting condition. 
  Figure \ref{chiq} shows  the $\mib{q}$ dependence of 
    $I_{\gamma\gamma'}(\mib{q})$ with $\mib{q} =(q_x, q_y)$
       (no $q_z$ dependence due to $t_z = 0$).
Since there are 
 maxima at $\mib{q}=\pm2\mib{k}_0=(\mp0.060\pi, \pm0.432\pi)$, 
  the  incommensurate density wave with  $\mib{q}=\pm2\mib{k}_0$ 
  is expected from $I_{12}(\mib{q})$ and $I_{21}(\mib{q})$, i.e., 
   the main contribution is given by  
     an excitation from the valence band ($\gamma =2$)
    to  the conduction  band ($\gamma =1$) 
       close to $\vep_F$ around $\mib{k}_0$. 

For the clear understanding of the effect of temperature on 
 the 2$\mib{k}_0$ density response, 
 we calculate  the susceptibility defined as 
%%%%%%%%%%%% (22) %%%%%%%%%%%%%%%%%%%%%%%%%%
\begin{align}
\chi^{0}_{\gamma\gamma'}(\mib{q})&=
-\dfrac{1}{N}\sum_{\mib{k}}
\dfrac{f(\xi_{\gamma}(\mib{k}+\mib{q}))-
f(\xi_{\gamma'}(\mib{k}))}
{\xi_{\gamma}(\mib{k}+\mib{q})-\xi_{\gamma'}(\mib{k})}
.
\label{susceptibility}
\end{align}
%%%%%%%%%%%%%%%%%%%%%%%%%%%%%%%%%%%%%%
 The temperature dependence of $\chi^{0}_{\gamma\gamma'}(2\mib{k}_0)$ is 
 shown in  Fig.~\ref{chi0}. 
With decreasing temperature, 
 the diagonal component, $\chi^{0}_{\gamma\gamma}(2\mib{k}_0)$, 
  corresponding to the intra-band process, 
   decreases and reduces to zero in the limit of zero temperature
    due to the vanishing of the density of states at the Fermi energy. 
However, 
 the off-diagonal component of $\chi^{0}_{12}(2\mib{k}_0) 
  (=\chi^{0}_{21}(2\mib{k}_0))$, 
    corresponding to the inter-band process rather increases 
      with decreasing temperature. 
The enhancement of the off-diagonal one 
  comes from the fact that the electron-hole excitation across the ZGS 
   can satisfy  the nesting condition.
  Note that, at the zero temperature,  
    the off-diagonal one does not diverge but has  a finite value.
   This  is ascribed to the ZGS 
      with  the vanishing of the density of state on the Fermi energy.  
 Thus it is found that 
   the 2$\mib{k}_0$ density wave is mainly determined 
     by the inter-band pairing.

%%%%%%%%%%%%%%%%%%%%%%%%%%%%%%%%%%%%%%
\begin{figure}[tbp]
\begin{center}
\includegraphics[width=7.0cm]{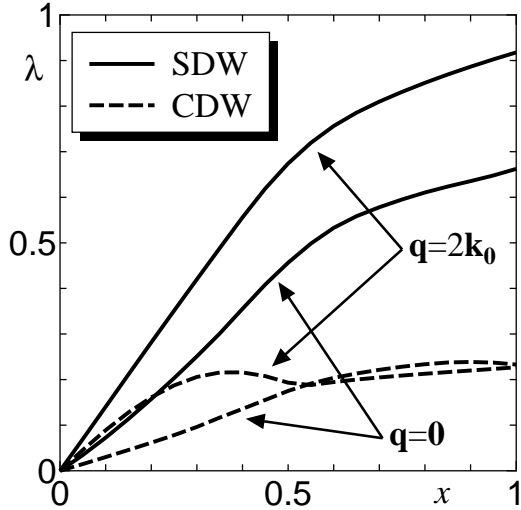}
\caption{$x$-dependence of $\lambda$ for the 
SDW (solid line) and CDW (dashed line) with $\mib{q}=0,2\mib{k}_0$, where
$p=16$ and $T=0.001$. 
The strength of the Coulomb interaction $x$ is defined as 
$U=0.4x, V_c=0.17x, V_p=0.05x$. 
}
\label{x-lambda}
\end{center}
\end{figure} 
%%%%%%%%%%%%%%%%%%%%%%%%%%%%%%%%%%%%%%

Here we examine the effect of interaction on the density wave
by calculating $\lambda$ in eqs.~(\ref{deltaxy}) and (\ref{deltac}). 
In addition to the direct  effect of interaction by $U$ and $V$, 
 there is also the effect through  $I_{\gamma\gamma'}$, which contains 
      $\lrangle{n_{\alpha}}$ in $\xi_{\gamma}$. 
  The variation of interactions are shown in 
   Fig.~\ref{x-lambda}, which denotes 
    $x$-dependence of   $\lambda$ 
        with $U=0.4x, V_c=0.17x, V_p=0.05x$ ($0 \leq x \leq  1$).
    The set of interactions for $x=1$\cite{kobayashi_2005_sc} 
   corresponds  to    the  parameters, which are taken to explain  
  the experimental result\cite{tajima_2002}.
  We note that, for $t_z = 0$, 
    such a choice of interactions ($x=1$) gives $T_C=0$ 
     due to  $\lambda < 1$
   while the further increase of  $x (> 1)$ leads to 
    a finite $T_C$ with  $\lambda = 1$. 
%   even for two-dimensional case. 
   However such a $T_C$  may be reduced to zero for $t_z=0$
     by the  two-dimensional fluctuation. 
  Within the present mean-field treatment, 
   the increase of $T_C$ is noticeable in the presence of 
    the interlayer transfer $t_z$ as  shown later. 

   In  Figs.~\ref{chi0} and \ref{x-lambda}, $\lambda$ for 
     the SDW and CDW states 
   %  with $\mib{q}=0,2\mib{k}_0$ 
   is calculated with 
    the transfer energies  at for $p=16$
    which is a reasonable pressure for the ZGS in the experiment.
    \cite{tajima_2006} 
%------------------------ 
The fact that $\lambda \propto x$ for small $\lambda$ is understood 
as follows. 
The main effect of interaction comes from the coefficient of 
r.h.s. of eqs.~(\ref{deltaxy}) and (\ref{deltac}) while 
the effect on $\lrangle{n_{\alpha}}$ is negligibly small. 
With increasing $x$ in all the curve, the $x$-linear dependence is 
suppressed indicating that the enhancement of the charge disproportionation 
by interactions suppresses 
 the density wave with $\mib{q}=0$ and $2\mib{k}_0$.
The reason for such a behavior
 partly comes from the suppression of the density of states close to 
 the Fermi surface by the increase of the  charge disproportionation. 
 The density response with  $\mib{q}=2\mib{k}_0$ 
   is larger than that with $\mib{q}=0$.   
  The increase for CDW as the function of $x$ 
  is due to the effect of  $V$. 

Although the $2\mib{k}_0$-SDW state is the largest one
    among these  four states, 
    the magnitude of the interaction  is not enough 
    to obtain the finite $T_C$ i.e. $\lambda = 1$ even for $x=1$
     in the present choice of parameters.

%--------------------------------------
\subsection{spin density wave in the presence of $t_z$}

Based on the results of the previous subsection, 
 we examine the SDW state,  which leads to 
 the finite $T_C$ in the presence of $t_z$.

%%%%%%%%%%%% Fig.7   %%%%%%%%%%%%%%%%%%%%%%%%%%
\begin{figure}[tbp]
\begin{center}
\includegraphics[width=8.5cm]{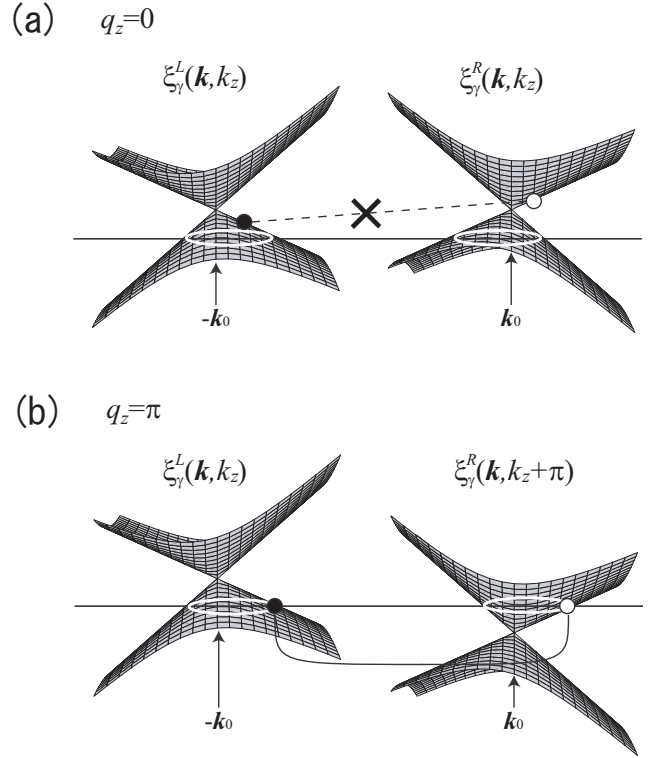}
\caption{Band dispersions $\xi^{R,L}_{\gamma}(\mib{k},k_z)$ $(\gamma=1,2)$ 
around $\mib{k} = \mib{k}_0$ and $- \mib{k}_0$ in the presence of $t_z$
for (a): $q_z=0$  and for (b): $q_z=\pi$, respectively.
 The Fermi surfaces are  drawn by the white circle.  
}
\label{fig_cone}
\end{center}
\end{figure} 
%%%%%%%%%%%%%%%%%%%%%%%%%%%%%%%%%%%%%%

%%%%%%%%%%%%%  Fig. 6   %%%%%%%%%%%%%%%%%%%%%%%%
\begin{figure}[tbp]
\begin{center}
\includegraphics[width=8.0cm]{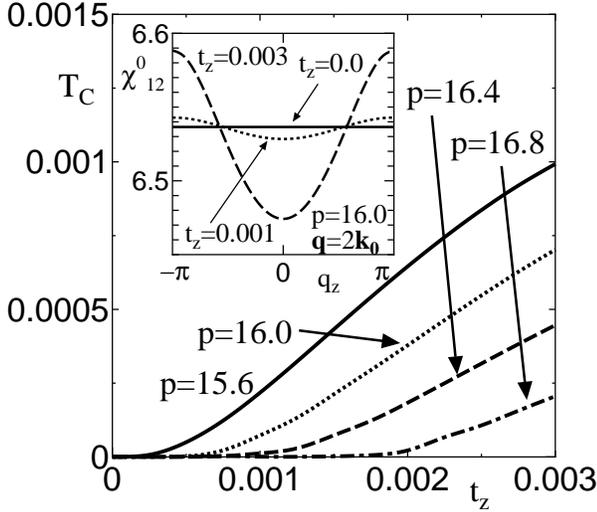}
\caption{
$t_z$-dependences of $T_C$ for $2\mib{k}_0$-SDW state with $q_z=\pi$
for several pressures of 
$p=15.6$ (solid line), $16.0$ (dotted line), $16.4$ (dashed line) 
and $16.8$ (dot dashed line), 
where Coulomb interactions are chosen as $U=0.4,V_c=0.17,V_p=0.05$. 
The inset figure is $q_z$-dependences of the inter-band 
susceptibility $\chi^0_{12}(2\mib{k}_0,q_z)$
given as eq. (\ref{susceptibility}) for 
$t_z = 0.0$ (solid line),  $0.001$ (dotted line) and $0.003$ (dashed line) at $p=16$, 
where the temperature is set as $T=0.001$. 
}
\label{tc-h6}
\end{center}
\end{figure} 
%%%%%%%%%%%%%%%%%%%%%%%%%%%%%%%%%%%%%%

%%%%%%%%%%%% Fig.8   %%%%%%%%%%%%%%%%%%%%%%%%%%
\begin{figure}[tbp]
\begin{center}
\includegraphics[width=7.5cm]{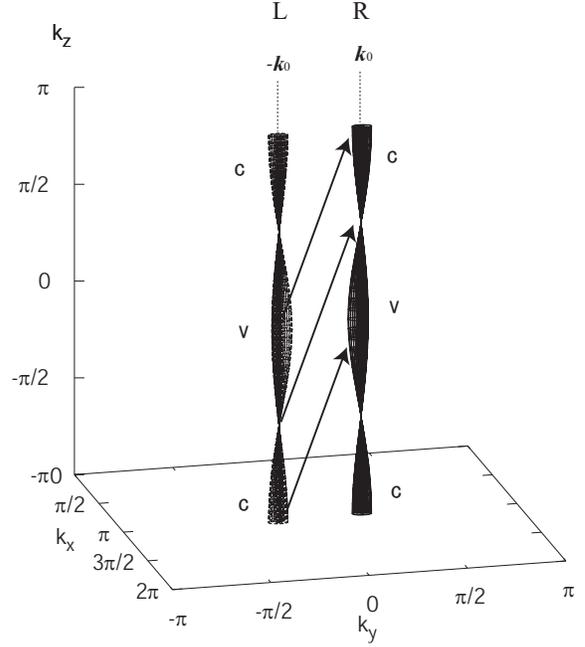}
\caption{Fermi surface in the three-dimensional momentum space, 
$(k_x, k_y, k_z)$, with  $k_z$ being the momentum along the interplane 
 where $U=0.4, V_c=0.17, V_p=0.05, p=16$ and $t_z=0.002$. 
The left one (L) and the right one (L) denote the Fermi surfaces  
around $-\mib{k}_0=(-0.970\pi,-0.216\pi)$ and  $\mib{k}_0$, respectively, 
( for convenience, $- \mib{k}_0$ is replaced by that of the extended zone). 
%$k_x\rightarrow k_x+2\pi$ for convenience). 
 The Fermi surface is given 
 by the valence band (v)
 %for $-\pi/2<k_z<\pi/2$  (we noted "v" in the figure), 
or  by the conduction band (c), 
%for $\pi/2<k_z<\pi$ and $\pi/2<k_z<\pi$ (we noted "c" in the figure), 
respectively, where 
the arrow denotes the nesting vector. 
}
\label{fig_3dfs}
\end{center}
\end{figure} 
%%%%%%%%%%%%%%%%%%%%%%%%%%%%%%%%%%%%%%

We note that $T_C$ takes a maximum at $q_z=\pi$,
 i.e., the SDW with the interlayer variation being out of phase. 
This fact  has been found  in the quasi-one-dimensional 
 density wave system, which consists of  an array of chains coupled with 
 the interchain electron hopping.
 \cite{Suzumura1978}
The typical $q_z$  dependence  of the present case appears in 
   the main contribution of the inter-band susceptibility 
   $\chi^0_{12}(2\mib{k}_0,q_z)$. 
 The inset of Fig.~\ref{tc-h6} displays 
   $\chi^0_{12}(2\mib{k}_0,q_z)$, which  takes a 
      maximum at $q_z =\pi$ and a minimum  at $q_z = 0$.
    The $q_z$ dependence of the susceptibility %$\chi^0_{12}(2\mib{k}_0)$ 
    is essentially given by $\chi^0_{12}(2\mib{k}_0,q_z) \propto \cos q_z$
     due to the dispersion 
       with $  2 t_z \cos k_z$ in eq.~(\ref{eigenvalue_kin}). 
       Here, we note  a fact that  $\chi^0_{12}(2\mib{k}_0,q_z)$ takes 
        a  maximum at  $q_z=\pm \pi$. 
   %     From  Figs. \ref{fig_cone}. 
   In  Fig.~\ref{fig_cone}, a typical example of the band dispersion around the     two contact points $\pm \mib{k}_0$ is shown, 
         where  $\xi_{\gamma}^{R}$ 
           and $\xi_{\gamma}^{L}$ denote 
          $\xi_{\gamma}(\mib{k},k_z)$ ($\gamma$ = 1,2) for 
           $\mib{k} \simeq  \mib{k}_0$  and  $\mib{k} \simeq - \mib{k}_0$,
            respectively.  
        For $q_z=0$, as shown in Fig. \ref{fig_cone} (a), 
      there is 
      no contribution from 
        the state with the wave number inside the circle of 
                the Fermi surface, 
     since electrons having $\xi_{\gamma}^{L}$ and 
       $\xi_{\gamma}^{R}$ are both occupied or vacant, 
       i.e. the inter-band process is absent. 
      For $q_z=\pi$, on the other hand,   
       the nesting condition is satisfied as shown in Fig. \ref{fig_cone} (b),
   for the valence  band of  $\xi_{\gamma}^{L}$ and the electron band of $\xi_{\gamma}^{R}$. 
     As for the state  outside of the circle of the Fermi surface, 
    it is also found that 
     the contribution of Fig. \ref{fig_cone} (b) is larger than 
     that of Fig. \ref{fig_cone} (a).
   Thus, the inter-band process is enhanced by   the appearance of two 
       Fermi surfaces given by  $\xi^{L}_{1(2)}$  with $k_z$ and 
         $\xi^{R}_{2(1)}$  with $k_z+\pi$,  
 %      The r.h.s of eq. (\label{deltaxy}) increases by the effect of $t_z$ 
 %      for $q_z=\pi$ and 
       and the SDW is optimized at $q_z=\pi$.

%---------------------- 
In Fig.~\ref{tc-h6} 
 the $t_z$ dependence of $T_C$ of  the 
$2\mib{k}_0$-SDW state is shown with some choices of pressures 
for $U=0.4, V_c=0.17, V_p=0.05$. 
 With increasing $t_z$, $T_C$ increases noticeably 
  while  $T_C$  for $t_z=0$ is negligibly small. 
The increase of $T_C$ for $q_z = \pi$ by $t_z$ 
 is ascribed to the grow of the Fermi surface of both
 valence and conduction bands, 
 as drawn in Fig. \ref{fig_3dfs},
 where  the nesting condition is partly kept 
  but is partly violated by $t_z$
  (e.g., there are two choices of pairings  in Fig. \ref{fig_cone} (b)). 
Note that $q_z = \pi$ is the relative momentum between 
$\xi_{\gamma}^{R}$      and $\xi_{\gamma}^{L}$ 
 and then does not depend on the sign of $t_z$.   
  The suppression of the SDW for $q_z = 0$ 
 comes from a fact that the region of the  Fermi surface participating in 
 the nesting condition  is always reduced  by $t_z$
  in spite of the emergence  the Fermi surface. 
  With increasing pressure, $T_C$ is reduced rapidly 
  due to the property of $p$-dependence of transfer energies, which  
   increase  the dip of the density of states 
    around the Fermi energy under hydrostatic pressures, or 
     the increase of the Fermi velocity under pressures.
   
%This originates from the reduction of the density of states around 
% Fermi energy due to 
% the increase of the Fermi velocity under pressures.

%%%%%%%%%%%% Fig. 7  %%%%%%%%%%%%%%%%%%%%%%%%%%
\begin{figure}[tbp]
\begin{center}
\includegraphics[width=8.5cm]{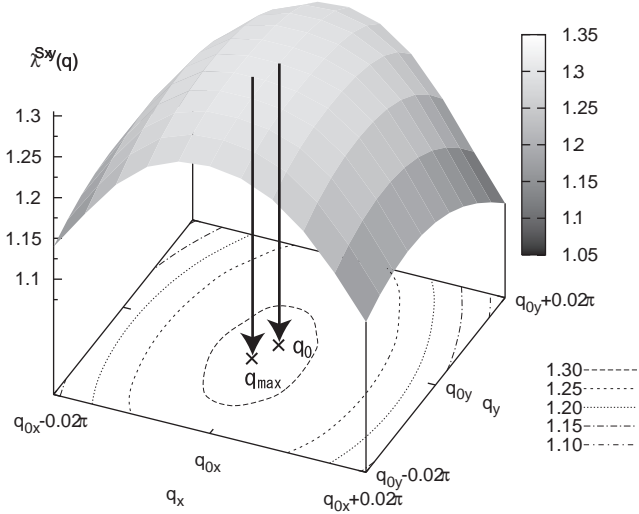}
\caption{$\mib{q}$-dependence of $\lambda^{S_{x,y}}$ in the presence of 
the Zeeman energy. Parameters are set as 
$U=0.4, V_c=0.17, V_p=0.05, p=16, H_0=0.003$ and $T=0.001$, where  
 $\mib{q}_0 =  2\mib{k}_0 = (-0.060\pi,0.432\pi)$ and 
 $\mib{q}_{max}=(-0.058\pi, 0.428\pi) $, respectively. 
}
\label{tc-h}
\end{center}
\end{figure} 
%%%%%%%%%%%%%%%%%%%%%%%%%%%%%%%%%%%%%%

\section{Summary and Discussion}

In summary, we have examined the role of 
interplane transfer energy on the  SDW state in the 
ZGS of the \eti salt under the hydrostatic pressure. 
First, we estimated  the pressure dependence of transfer energy.
% and charge disproportionation{\bf (this also must be described) }. 
The ZGS appears at pressures higher than those  of 
uniaxial strain along  the stacking axis. 
Our result is qualitatively  consistent with the experiment. 
Next, we studied the SDW in the presence of $t_z$
 and showed the increase of  
$T_C$ with increasing $t_z$. 
Since the SDW originates from the electron-hole  excitons between  two Dirac cones
 with $\pm \mib{k}_0$,
  the nesting condition is reduced due to the  Fermi point 
   for the two-dimensional case of $t_z=0.$ 
Thus the appearance  of the Fermi surface by  $t_z$ gives rise to 
 the SDW. 

Here we mention about  the location of $\mib{q} = \mib{q}_{max}$, which gives 
 the  maximum value of $T_C$.
 Within the numerical accuracy of the present calculation,
 $\mib{q}_{max}$ coincides with  $2\mib{k}_0$ even  for $t_z \not= 0$  
 although the deviation is generally expected.  
 Actually  $\mib{q}_{max}$ becomes slightly  
 different from   $2\mib{k}_0$ for the case of the Zeeman energy,
$ - H_0 \sum_{i \alpha \sigma} {\rm sgn} (\sigma) 
 a_{i \alpha \sigma}^{\dagger}a_{i \alpha \sigma}$,  
 instead of $t_z$  ($\mu_B=1$), 
where $\mathrm{sgn}(\sigma)=1(-1)$ for 
$\sigma=\uparrow(\downarrow)$. 
 In the presence of $H_0$,  
 $\lambda^{S_{x,y}}$ increases but $\lambda^{S_{z}}$ decreases. 
Note that the effect of $H_0$ on $T_C$ of $S_{x,y}$ is 
 nearly the same but is slightly large compared with $t_z$.
Figure \ref{tc-h} shows an example of    
 the contour plot of  $\lambda$ of $S_{x,y}$
on the plane of 
 $\mib{q} = (q_x,q_y)$ close to $\mib{q}=2\mib{k}_0(=(-0.060\pi, 0.432\pi))$, 
where the parameters 
are chosen as $U=0.4,V_c=0.17,V_p=0.05$ 
%and $t_z= \cdots$
with  $p=16$ and  $H_0 =0.003$. 
 The location of the maximum is shown by the cross.
Although the valence and conduction bands  coincide  at $\pm\mib{k}_0$, 
$\mib{q}_{max}(=(-0.058\pi, 0.428\pi))$  is slightly different from $2\mib{k}_0$. 
%the maximum value of $\lambda$ is given by 
%$\mib{q} = \mib{q}_{max}$, which is different from $2\mib{k}_0$. 
%Figure shows the Fermi surface of the \eti salt, which 
%appear by the magnetic field near $\pm\mib{k}_0$.
% {\bf (where is the corresponding figure?) }
Since the Dirac cone is anisotropic, 
the center of the Fermi surface with an ellipse is not located 
on $\mib{k}_0$. 
Thus  we can see
% $\mib{q}=2\mib{k}_0$ does not give the optimum condition for the nesting
% for the present salt.
 the optimum wave number is near $\mib{q}=2\mib{k}_0$ but 
 is not exactly the same. 
 On the other hand, $\mib{q}_{max}$ for $t_z \not= 0$
 is less effected since $t_z \cos k_z$ gives  contribution 
  corresponding to both  $\pm H_0$. 
% the cobtirution does not change 
% by increasing $t_z$. 
% since the linear dispersion is anisotropic, 
%the nesting vector is not equal to $\mib{q}=2\mib{k}_0$.

We discuss the estimation of transfer energies and ZGS 
 under the hydrostatic pressure. 
%In  the calculations of the energy  band 
%under uniaxial pressure along $a$-axis, $P_a$, 
%the ZGS appears for  $p>3$ when $U=V_p=V_c=0$ and 
%for  $p>4.3$ when $U=0.4, V_p=0.05, V_c=0.17$. 
As shown in the previous section,
the ZGS state under the hydrostatic pressure emerges at higher 
pressure 
compared  with ZGS  under uniaxial pressure  $p_a$. 
This result is consistent with the experiment, 
since the temperature-independent behavior indicating the ZGS 
is observed at $p_a=10$ (uniaxial pressure) and at $p=20$ (hydrostatic pressure). 
However our result for ZGS state is still under  smaller pressure. 
%In the \eti salt 
At high pressures with  $p>10$\cite{tamura}, 
 it is expected  that the distortion is suppressed 
    due to  the deviation from the  Hooke's law.  
Thus the increase of $t_A$ by the pressure could be suppressed 
for that region and the ZGS is realized under higher pressure than 
our estimated one.

Finally, we note on the interlayer hopping. 
For simplicity, $t_z$ is taken for the hopping, 
in which the electron moves into a site with the same index $\alpha(=1,\cdots,4)$ of the adjacent layer, 
and then the position of the contact point is independent of $k_z$ 
(see eq. (\ref{eigenvalue_kin})). 
 However    the hopping into other sites, which may reduce 
 the nesting condition, is expected from 
     the $k_z$ dependence of the contact point as  obtained  
   by the first principle calculation.\cite{kino_1}
 For the moment,  there is no data of  extended H\"uckel method 
   for interlayer hoppings,  and  
%  calculated by  in terms of , 
 the role of the three-dimensionality remains as a future problem 
  to clarify the properties of ZGS in \eti salt.

%Although organic conductors are softer than conventional metals, 
%the calculated Poisson's ratio ($\nu\sim0.1$) is too small compared 
%to the conventional one ($\nu\sim0.3$). 
%When the Poisson's ratio is larger, 
%coefficients 
%between transfer energies and pressure becomes smaller 
%leading to  the ZGS at higher pressures.{\bf (must be checked) } 

\vspace{1cm}

\section*{Acknowledgements}
We are grateful to N. Tajima, R. Kondo and H. Fukuyama 
for useful discussions. 
The authors  also thank M. Tokumoto for informing us the reference
 concerning the hydrostatic pressure effect.\cite{tamura}  
S. K. acknowledges the financial support of Research Fellowship for 
Young Scientists from Japan Society for the Promotion of Science (JSPS). 
This work also financially supported by a Grant-in-Aid for Scientific Research on 
Priority Areas of Molecular Conductors (No. 15073103) from the Ministry of Education, 
Culture, Sports, Science and Technology, Japan.

\vspace{1cm}

%------------------------
\appendix
\section{Estimation of $r_1$ and $r_2$}
%-----------------------------
We calculate quantities $r_1$ and $r_2$ 
 from the data of the lattice parameters under 
the uniaxial strain\cite{kondo} and hydrostatic pressure\cite{tamura}. 
The variation of the lattice constants,
 which are  defined as $u_a=(a(p_a,p_b)-a_0)/a_0$ for $a$-axis 
and $u_b=(b(p_a,p_b)-b_0)/b_0$ for $b$-axis, 
  is written as 
%%%%%%%%%%%% (A1) %%%%%%%%%%%%%%%%%%%%%%%%%
\begin{align}
\left(
\begin{array}{c}
u_a\\
u_b
\end{array}
\right)
&=-
\left(
\begin{array}{cc}
s_1&s_2\\
s_3&s_4
\end{array}
\right)
\left(
\begin{array}{c}
p_a\\
p_b
\end{array}
\right) ,  
\label{uaub}
\end{align}
%%%%%%%%%%%%%%%%%%%%%%%%%%%%%%%%%%%%%%
where $a_0=a(0,0)$ and $b_0=b(0,0)$. 
From eqs.~(\ref{ratio_P}) and (\ref{uaub}), 
% lattice 
parameters 
 $r_1, r_2, s_1, s_2, s_3$ and $s_4$
are related with the lattice distortion per unit pressure, $E_j$,  
$(j=1, \cdots 6)$. Actually we obtain 
%%%%%%%%%%%%% (A2) (A3) (A4)%%%%%%%%%%%%%%%%%%%%%%%%%
\begin{align}
s_1+s_2=E_{1},&&
s_3+s_4=E_{2},
\label{lattice_equation1}\\
s_1+s_2r_1=E_{3},&&
s_3+s_4r_1=E_{4},
\label{lattice_equation2}\\
s_1r_2+s_2=E_{5},&&
s_3r_2+s_4=E_{6},
\label{lattice_equation3}
\end{align}
%%%%%%%%%%%%%%%%%%%%%%%%%%%%%%%%%%%%%%
where eqs. (\ref{lattice_equation1}), (\ref{lattice_equation2}) and 
(\ref{lattice_equation3}) 
correspond to  the hydrostatic pressure, $a$-axis strain and $b$-axis strain, 
respectively. 
From the experimental results of the lattice distortion,
\cite{tamura,kondo} 
$E_i\ (i=1,\cdots,6)$ are given  as 
$E_1=0.00265\ {\rm kbar}^{-1}$, $E_2=0.00247\ {\rm kbar}^{-1}$, $E_3=0.00326\ {\rm kbar}^{-1}$, 
$E_4=0.0\ {\rm kbar}^{-1}$, $E_5=0.0\ {\rm kbar}^{-1}$ and $E_6=0.00340\ {\rm kbar}^{-1}$. 
By solving eqs. (\ref{lattice_equation1}), (\ref{lattice_equation2}) and 
(\ref{lattice_equation3}),
 we obtain 
$ s_1 = 0.00357\ {\rm kbar}^{-1}$, 
$ s_2 = - 0.000917\ {\rm kbar}^{-1}$, 
$ s_3 = - 0.00124\ {\rm kbar}^{-1}$, 
$ s_4 = 0.00371\ {\rm kbar}^{-1}$, 
$r_1 = 0.335$ and 
$r_2 = 0.257$.
% for $r_1,r_2,s_1,\cdots,s_4$, 
%$K^a_A$ and $K^b_A$ can be calculated and the 
%proportional coefficients of transfer energies for the 
%hydrostatic pressure, 
%Substituting $r_1$ and $r_2$ into eq.~(\ref{eq:K-L}),
% $K^a_A+K^b_A$ are calculated as shown in Table \ref{table_coefficient}. 
 
 We note that quantities $r_1, \cdots , s_4$ can be expressed 
 in terms of  Young's modulus, $Y_a$ and $Y_b$, and 
 Poisson's ratio, $\nu_{ab}$ and $\nu_{ba}$, which are  
  defined as 
%%%%%%%%%%% (A5) (A6) %%%%%%%%%%%%%%%%%%%%%%%%%%%
\begin{align}
Y_a=\left(\dfrac{p_a}{u_a}\right)_{p_b=0}&,& 
Y_b=\left(\dfrac{p_b}{u_b}\right)_{p_a=0}\\
\nu_{ab}=-\left(\dfrac{u_a}{u_b}\right)_{p_a=0}&,& 
\nu_{ba}=-\left(\dfrac{u_b}{u_a}\right)_{p_b=0}.
\end{align}
%%%%%%%%%%%%%%%%%%%%%%%%%%%%%%%%%%%%%%
Since
$s_1 = 1/Y_a$, 
$s_2 = -r_2/Y_a$, 
$s_3 = -r_1/Y_b$, 
$s_4 = 1/Y_b$, 
we obtain 
$ r_1 = \nu_{ba} Y_b/Y_a$ and 
$ r_2 = \nu_{ab} Y_a/Y_b$.

\end{document}